\documentclass[aps]{revtex4}
\usepackage{graphicx}
\begin{document}
%
%
\title
  {\bf Isospin mixing within the multi-reference nuclear density functional theory and beyond - selected aspects
  }
  \altaffiliation
     {This work was supported in part by the Polish National Science Center
and by the Academy of Finland and
University of Jyv\"askyl\"a within the FIDIPRO programme. We
acknowledge the CSC - IT Center for Science Ltd, Finland, for the
allocation of computational resources.
     }
%
%
\author{W.~Satu{\l}a$^{1}$}
\author{J.~Dobaczewski$^{1,2}$}
\author{M.~Konieczka$^{1}$}
\author{W.~Nazarewicz$^{1,3,4}$}

%
%
\affiliation{$^{1}$Faculty of Physics, University of Warsaw, 
PL-00-681 Warsaw, Poland}
\affiliation{$^{2}$Department of Physics, 
University of Jyv\"askyl\"a, FI-40014  Jyv\"askyl\"a, Finland}
\affiliation{$^{3}$Department of Physics and
Astronomy, University of Tennessee, Knoxville, Tennessee 37996, USA}
\affiliation{$^{4}$Physics Division, Oak Ridge National Laboratory, Oak Ridge, Tennessee 37831, USA}

\begin{abstract}
The results of systematic calculations of isospin-symmetry-breaking
corrections to  superallowed $\beta$-decays based on the self-consistent isospin- and  angular-momentum-projected
nuclear density functional theory (DFT) are reviewed with an emphasis on theoretical uncertainties
of the model. Extensions of the formalism
towards {\it no core\/} shell model approach with basis cutoff scheme dictated by the self-consistent
particle-hole DFT solutions will be also discussed.
\end{abstract}

%
\maketitle
%
%

\section{Introduction}

Isospin symmetry in atomic nuclei is weakly broken mostly by the Coulomb interaction
that exerts a long-range polarization effect. Capturing an equilibrium between
long and short range effects is a challenging task possible only within {\it no core\/} approaches,
which, in heavier nuclei, reduces possible choices
to formalisms rooted in the density functional theory (DFT). However, as it was recognized already in
the 70's~\cite{[Eng70]}, the self-consistent mean-field (MF)
approaches cannot be directly applied to compute isospin impurities because of spurious
mixing caused by the spontaneous symmetry breaking (SSB) effects. This observation hindered
theory from progress in the field for decades.

The aim of this work is to present a brief overview of recent theoretical results obtained
within the isospin- and angular-momentum projected DFT on isospin-mixing
effects. Our multi-reference {\it no core\/} DFT
was specifically designed to treat rigorously the conserved rotational symmetry and, at the
same time, tackle the explicit breaking of the isospin symmetry due to the Coulomb field.
The major physics motivation behind developing the model and studying the isospin symmetry breaking (ISB)
comes from nuclear beta decay. Theoretical corrections to the superallowed Fermi beta decay
matrix elements $I=0^+,T=1\rightarrow I=0^+,T=1$ between the  isobaric analogue states,
caused by the ISB,  are critical for precise
determination of the leading  element $V_{ud}$ of the Cabibbo-Kobayashi-Maskawa (CKM)
flavour-mixing matrix and, in turn, for further stringent tests of its unitarity,
violation of which may signalize {\it new physics\/}
beyond the Standard Model of particle physics, see~\cite{[Tow10a]} and refs. quoted therein.

\section{MULTI-REFERENCE DENSITY FUNCTIONAL THEORY}

The formalism employed here starts with the self-consistent Slater determinant
$|\varphi \rangle$ obtained by solving Skyrme-Hartree-Fock equations without
pairing. The state violates both the rotational and isospin symmetries. The strategy
is to restore the rotational invariance, remove the spurious isospin mixing caused
by the isospin SSB effect, and retain only the physical isospin mixing
due to the electrostatic interaction~\cite{[Sat09sx]}.  This is
achieved by a rediagonalization of the entire Hamiltonian,
consisting the isospin-invariant kinetic energy and Skyrme force and
the isospin-non-invariant Coulomb force, in a basis that conserves both angular
momentum and isospin $|\varphi ;\, IMK;\, TT_z\rangle$,
projected from the state $|\varphi \rangle$:
\begin{equation}\label{ITbasis}
|\varphi ;\, IMK;\, TT_z\rangle =   \frac{1}{\sqrt{N_{\varphi;IMK;TT_z}}}
\hat P^T_{T_z,T_z} \hat P^I_{M,K} |\varphi \rangle ,
\end{equation}
where $\hat P^T_{T_z ,T_z}$ and $\hat P^I_{M,K}$ stand for the standard isospin
and angular-momentum projection operators~\cite{[RS80]}, respectively.
One must also treat the fact that the quantum number
$K$ is not conserved and set (\ref{ITbasis}) is overcomplete.
This requires selecting the subset of linearly independent states,
known  as {\it collective space}~\cite{[RS80]}, which
is spanned, for each $I$ and $T$, by the so-called {\it natural states\/}
$|\varphi;\, IM;\, TT_z\rangle^{(i)}$ \cite{[Dob09ds]} and subsequently
rediagonalizing the entire Hamiltonian in the collective space. The resulting
eigenfunctions are:
\begin{equation}\label{KTmix}
|n; \,\varphi ; \,
IM; \, T_z\rangle =  \sum_{i,T\geq |T_z|}
   a^{(n;\varphi)}_{iIT} |\varphi;\, IM; TT_z\rangle^{(i)} ,
\end{equation}
where index $n$ labels the eigenstates in ascending order
of energies.

\section{ISOSPIN MIXING}

The isospin or Coulomb impurities are defined as:
\begin{equation}
\label{truemix}
\alpha_{\rm C}^n = 1 - \sum_{i} |a^{(n;\varphi)}_{iIT}|^2,
\end{equation}
where the sum extends over the norms
corresponding to isospin $T$ that dominates in wave function (\ref{KTmix}).

\begin{figure}\begin{center}
\includegraphics[angle=0,width=0.60\textwidth,clip]{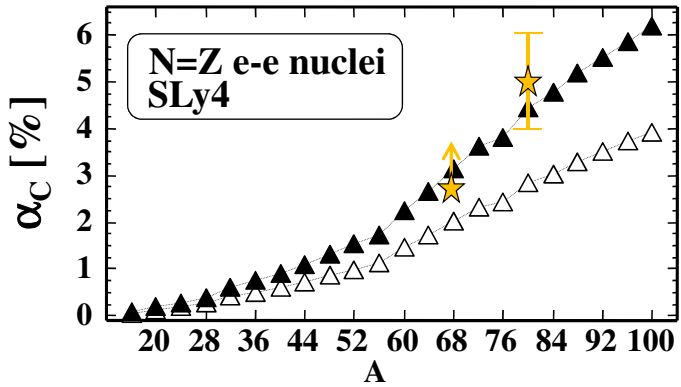}\hspace{0.05\textwidth}%
\begin{minipage}[b]{0.25\textwidth}
\caption[T]{\label{mixing}
Isospin impurities in even-even $N=Z$ nuclei calculated
by using the SLy4 Skyrme EDF~\protect\cite{[Cha97w]}. Full triangles mark the values calculated
by using the isospin-projected DFT. Open triangles show mean-field values that are artificially
quenched by the spurious isospin mixing. Stars mark empirical results in
$^{64}$Ge~\protect\cite{[Far03]} and $^{80}$Zr~\protect\cite{[Cor11x]}.}
\end{minipage}
\end{center}\end{figure}

It is well known that for modern density-dependent Skyrme and
Gogny energy density functionals (EDFs), the angular momentum projection is ill-defined~\cite{[Ang01],[Zdu07]}. Hence, at present,
the double-projected DFT method can be safely used only with the functionals originating from the true
Hamiltonian. Nevertheless, for all modern Skyrme forces, the isospin-only-projected variant of the approach is free
from singularities~\cite{[Sat09sx]}. Fig.~\ref{mixing} shows the isospin impurities
in the ground-states of even-even $N=Z$ nuclei, calculated by using the state-of-the-art SLy4 Skyrme~\cite{[Cha97w]} EDF
in the isospin-only-projected variant of the model~\cite{[Sat09sx],[Sat11sx]}.
It is gratifying to see that the calculated impurities are consistent with the recent data extracted
from the giant-dipole-resonance decay studies in $^{80}$Zr~\cite{[Cor11x]} and isospin-forbidden E1 decay in
$^{64}$Ge~\cite{[Far03]}, see Fig.~\ref{mixing}. Both data points disagree with the pure MF results,
which, due to the spurious mixing caused by
the spontaneous ISB effects, are
lower by almost $\sim$30\%. The agreement with available data
indicates that the model is capable of quantitatively capturing the
intensity of the isospin mixing. This is important in the context of
performing reliable calculations of the ISB corrections to the Fermi
beta decay.

\section{ISOSPIN-SYMMETRY-BREAKING CORRECTIONS TO THE SUPERALLOWED FERMI BETA DECAY}

The $0^+  \rightarrow 0^+$ Fermi $\beta$-decay proceeds between
the ground state (g.s.) of the even-even nucleus
$|I=0, T\approx 1, T_z = \pm 1 \rangle$ and its isospin-analogue partner
in the $N=Z$ odd-odd nucleus, $| I=0, T\approx 1, T_z = 0 \rangle$.
Since the isospin projection alone leads to unphysically large isospin mixing
in odd-odd $N=Z$ nuclei~\cite{[Sat11sx]}, to calculate Fermi matrix elements
the double-projected method
must be applied. As already mentioned, the
angular momentum projection brings
back the singularities in the energy kernels~\cite{[Sat11sx]}, preventing
one from using the modern parametrizations of the Skyrme EDFs and forcing
us to use the Hamiltonian-driven Skyrme SV EDF~\cite{[Bei75s]}.

The g.s.\
of the even-even parent nucleus is approximated by the state
$|\psi ;\, I=0, T\approx 1, T_z = \pm 1 \rangle$, projected from the Slater determinant
$|\psi \rangle$ representing the self-consistent g.s.\ MF solution,
which is unambiguously defined by filling in the pairwise
doubly degenerate levels of protons and neutrons up to the Fermi level.

The odd-odd daughter state is approximated by
the state $|\varphi ;\, I=0, T\approx 1, T_z = 0 \rangle$, projected from the self-consistent Slater determinant
$|\varphi \rangle \equiv |\bar \nu \otimes \pi \rangle$ (or  $| \nu \otimes
\bar \pi \rangle$) representing the so-called anti-aligned MF configuration,
obtained by placing the odd neutron and odd proton in
the lowest available time-reversed (or signature-reversed) single-particle orbits.
The isospin projection from Slater determinants manifestly breaking
the isospin symmetry is essentially the only way to reach the $T\approx 1$  states
in odd-odd $N=Z$ nuclei that are beyond the MF model space.

\begin{figure}[t]
\includegraphics[width=0.9\columnwidth,clip]{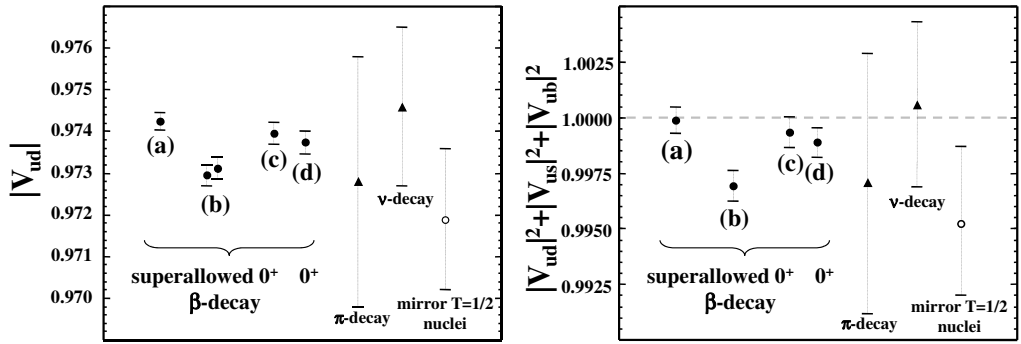}
\caption[T]{\label{fig2}
Matrix elements $|V_{\rm ud}|$ (left panel) deduced from the superallowed
$0^+\rightarrow 0^+$ $\beta$-decay (dots) by using values of $\delta_{\rm C}$ calculated in:
(a) Ref.~\cite{[Tow08]}; (b)  Ref.~\cite{[Lia09]} with NL3
and DD-ME2 Lagrangians; and in our work Ref.~\cite{[Sat12s]} with (c)  SV  and (d) SHZ2
EDFs. Triangles mark values obtained from the pion-decay \cite{[Poc04]} and
neutron-decay \cite{[Nak10]} studies. The open
circle shows the value of $|V_{\rm ud}|$  deduced from the $\beta$-decays in the
$T=1/2$ mirror nuclei \cite{[Nav09a]}.
Right panel shows the unitarity condition for different values of $|V_{\rm ud}|$
\cite{[Nak10]}.
}.
\end{figure}

This allows for rigorous fully quantal evaluation of the beta-decay
transition matrix element
and the corresponding ISB correction $\delta_{\rm C}$:
\begin{equation}\label{fermime}
|M_{\rm F}^{(\pm )}|^2 =
|\langle\psi ;\, I=0, T\approx 1,
T_z = \pm 1 | \hat T_{\pm} | \varphi ;\, I=0, T\approx 1, T_z = 0 \rangle |^2
\equiv  2(1-\delta_{\rm C}) .
\end{equation}
The calculated ISB corrections $\delta_{\rm C}$ lead to $|V_{ud}|=0.97397(27)$ and $|V_{ud}|=0.97374(27)$, for the
SV and SHZ2 EDFs, respectively~\cite{[Sat12s]}. Both values result in  the
unitarity of the CKM matrix  up to 0.1\%.
The new parametrization SHZ2 has been specifically developed to asses the robustness
of our results with respect to the choice of interaction.
This shows that although individual ISB corrections are sensitive to the interplay between the bulk
symmetry energy and time-odd mean-fields, the value of $|V_{ud}|$ rather weakly depends on the
parametrization. It is gratifying to see that our results are fully consistent  with the
results obtained by Towner and Hardy~\cite{[Tow08]}, which were obtained
within a  different methodology, based on the
nuclear shell-model combined with mean-field wave functions. Both approaches disagree with the RPA-based study
of Ref.~\cite{[Lia09]}. The theoretical results are summarized in Fig.~\ref{fig2}.


\section{OUTLOOK: BEYOND THE MULTI-REFERENCE DFT}

\begin{figure}\begin{center}
\includegraphics[angle=0,width=0.50\textwidth,clip]{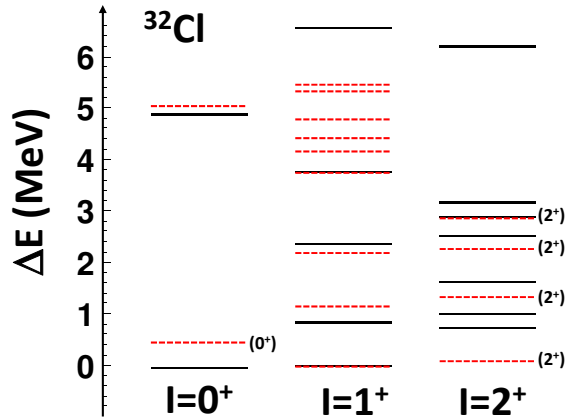}\hspace{0.05\textwidth}%
\begin{minipage}[b]{0.25\textwidth}
\caption[T]{\label{fig3}
Low-lying low-spin $I=0,1,2$ states in $^{32}$Cl. Solid lines
show theoretical levels obtained by mixing states projected from nine low-lying
self-consistent particle-hole configurations. Dashed lines mark experimental
data~\protect\cite{[Oue11]}. Note that spin assignments of, in particular, 2$^+$ states are uncertain.
The spectra have been normalized to the lowest $1^+$ state.
}
\end{minipage}
\end{center}\end{figure}

Implementation of the theory that we presented above was based on a projection from a single
Slater determinant, which, in odd-odd daughter nucleus, was not uniquely defined.
At present, we are implementing an extended version of the model, which allows
for mixing of states
projected from different self-consistent Slater determinants representing low-lying
(multi)particle-(multi)hole excitations in a given nucleus. Such an extension can be viewed
as a variant of {\it no core\/} shell-model with two-body effective interaction (including the
Coulomb force)  and a basis truncation scheme dictated by the self-consistent deformed Hartree-Fock solutions.
Preliminary spectrum of low-spin $I=0,1,2$ states in $^{32}$Cl obtained by mixing states projected from nine low-lying
particle-hole configurations is shown in Fig.~\ref{fig3}. In spite of certain technical problems
related to divergencies, which will be discussed elsewhere, the results are very encouraging.
This is particularly the case in view of the fact
that the self-consistent states and their mixing were
determined by using the SV Skyrme EDF, which has rather poor
spectroscopic properties.

%

%

\begin{thebibliography}{10}

\bibitem{[Eng70]}
{C.A. Engelbrecht and R.H. Lemmer, Phys. Rev. Lett. {\bf 24}, 607 (1970)}.

\bibitem{[Tow10a]}
{I.S. Towner and J.C. Hardy, Rep. Prog. Phys. {\bf 73}, 046301 (2010)}.

\bibitem{[Sat09sx]}
{W. Satu{\l}a {\it et al.}, Phys. Rev. Lett. {\bf 103}, 012502 (2009); Phys.
  Rev. C {\bf 81}, 054310 (2010)}.

\bibitem{[RS80]}
{P. Ring and P. Schuck, {\sl The Nuclear Many-Body Problem} (Springer-Verlag,
  Berlin, 1980)}.

\bibitem{[Dob09ds]}
{J. Dobaczewski {\it et al.}, Comput. Phys. Commun. {\bf 180}, 2361 (2009)}.

\bibitem{[Cha97w]}
{E. Chabanat, {\it et al.\/}, Nucl. Phys. {\bf A627} (1997) 710; {\bf A635}
  (1998) 231.}

\bibitem{[Far03]}
{E. Farnea {\it et al.}, Phys. Lett. {\bf B551}, 56 (2003)}.

\bibitem{[Cor11x]}
{A. Corsi {\it et al.}, Acta Phys. Pol. {\bf B42}, 619 (2011); Phys. Rev. C
  {\bf 84}, 041304 (2011)}.

\bibitem{[Ang01]}
{M. Anguiano, J.L. Egido, and L.M. Robledo, Nucl. Phys. {\bf A696}, 467
  (2001)}.

\bibitem{[Zdu07]}
{H. Zdu{\'n}czuk, J. Dobaczewski, and W. Satu{\l}a, Int. J. Mod. Phys. E {\bf
  16}, 377 (2007)}.

\bibitem{[Sat11sx]}
{W. Satu{\l}a {\it et al.}, Acta Phys. Pol. {\bf B42}, 415 (2011); Int. J. Mod.
  Phys. {\bf E20}, 244 (2011)}.

\bibitem{[Bei75s]}
{M. Beiner {\it et al.\/}, Nucl. Phys. {\bf A238}, 29 (1975)}.

\bibitem{[Tow08]}
{I.S. Towner and J.C. Hardy, Phys. Rev. C {\bf 77}, 025501 (2008)}.

\bibitem{[Lia09]}
{H. Liang, N. Van Giai, and J. Meng, Phys. Rev. C {\bf 79}, 064316 (2009)}.

\bibitem{[Sat12s]}
{W. Satu{\l}a {\it et al.}, Phys. Rev. C {\bf 86}, 054314 (2012)}.

\bibitem{[Poc04]}
{D. Pocanic {\it et al.}, Phys. Rev. Lett. {\bf 93}, 181803 (2004)}.

\bibitem{[Nak10]}
{K. Nakamura {\it et al.} (Particle Data Group), J. Phys. G {\bf 37}, 075021
  (2010)}.

\bibitem{[Nav09a]}
{O. Naviliat-Cuncic and N. Severijns, Phys. Rev. Lett. {\bf 102}, 142302
  (2009)}.

\bibitem{[Oue11]}
{C. Ouellet and B. Singh, Nuclear Data Sheets {\bf 112}, 2199 (2011)}.

\end{thebibliography}
%

\end{document}